# Symbol Detection in a MIMO Wireless Communication System Using a FeFET-coupled CMOS Ring Oscillator Array

**Harsh Kumar Jadia[1], Abhinaba Ghosh[2], Md Hanif Ali[1], Syed Farid Uddin[1], Sathish N[1], Shirshendu Mandal[1], Nihal Raut[1], Halid Mulaosmanovic[3], Stefan Dünkel[3], Sven Beyer[3], Suraj Amonkar[4], Udayan Ganguly[1,5], Veeresh Deshpande[1,5] & Debanjan Bhowmik[1,5]**

[1]Department of Electrical Engineering, Indian Institute of Technology Bombay, Mumbai, Maharashtra 400076, India

[2]Centre for Machine Intelligence and Data Science (CMInDS), Indian Institute of Technology Bombay, Mumbai, Maharashtra 400076, India

[3]GlobalFoundries Fab1 LLC and Co. KG, Dresden 01109, Germany

[4]Fractal AI Research, Fractal Analytics, Level 7, Commerz II International Business Park, Oberoi Garden City, Goregaon, Mumbai, Maharashtra 400063, India

[5]Centre for Semiconductor Technologies (SemiX), Indian Institute of Technology Bombay, Mumbai, Maharashtra 400076, India

*Correspondence should be addressed to DB (debanjan@ee.iitb.ac.in)*

**Abstract:** Symbol decoding in multiple-input multiple-output (MIMO) wireless communication systems requires the deployment of fast, energy-efficient computing hardware deployable at the edge. The brute-force, exact maximum likelihood (ML) decoder, solved on conventional classical digital hardware, has exponential time complexity. Approximate classical solvers implemented on the same hardware have polynomial time complexity at the best. In this article, we design an alternative ring-oscillator-based coupled oscillator array to act as an oscillator Ising machine (OIM) and heuristically solve the ML-based MIMO detection problem. Complementary metal oxide semiconductor (CMOS) technology is used to design the ring oscillators, and ferroelectric field effect transistor (FeFET) technology is chosen as the coupling element (X) between the oscillators in this CMOS + X OIM design. For this purpose, we experimentally report high linear range of conductance variation (1 micro-S to 60 micro-S) in a FeFET device fabricated at 28 nm high-K/ metal gate (HKMG) CMOS technology node. We incorporate the conductance modulation characteristic in SPICE simulation of the ring oscillators connected in an all-to-all fashion through a crossbar array of these FeFET devices. We show that the above range of conductance variation of the FeFET device is suitable to obtain optimum OIM performance with no significant performance drop up to a MIMO size of 100 transmitting and 100 receiving antennas, thereby making FeFET a suitable device for this application. Our simulations and associated analysis using the Kuramoto model of oscillators also predict that this designed classical analog OIM, if implemented experimentally, will offer logarithmic scaling of computation time with MIMO size, thereby offering a huge improvement (in terms of computation speed) over aforementioned MIMO decoders run on conventional digital hardware.

**Introduction**

Computing using coupled oscillator arrays forms a very interesting and useful sub-paradigm within the larger paradigm of physical or natural computing[1,2]. Coupled oscillator models have been used to explain various aspects of working of the brain including human memory formation[1,3-6]. Coupled oscillator arrays based on emerging hardware have also been proposed for data classification as an alternative to artificial neural network (ANN) algorithms implemented on both conventional digital and analog crossbar array hardware[7,8]. The coupling element between any two oscillators can be programmed much like synapses connecting biological neurons in the brain and artificial neurons in crossbar hardware are programmed through potentiation and depression[1]. Because of all the above reasons, coupled oscillator arrays are also known as oscillatory neural networks (ONN), and their design and performance are of major interest[1,9].

The Lyapunov function associated with a coupled oscillator array can be correlated with the energy landscape associated with the Ising model of spins in physics (Ising Hamiltonian)[2]. Hence, such an ONN has been proposed and experimentally demonstrated to act as a particular classical analog system known as the oscillator Ising machine (OIM). The OIM can solve various computationally challenging NP-Hard combinatorial optimization problems by minimizing the Lyapunov function (reaching the ground state or a close-to-ground state of the equivalent classical Ising Hamiltonian) just like other Ising machines can solve such problems following other techniques[10–19].

OIMs can be implemented through either purely conventional silicon complementary metal oxide semiconductor (CMOS)[9,12,20–24], or CMOS-compatible emerging devices (metal-insulator phase transition devices[25,26], spintronic devices [13,27–

[30]), or through a combination of CMOS and emerging devices like the ferroelectric field effect transistor (FeFET)[31–34]. Hardware of the third kind is also known as "CMOS + X" hardware[32,35,36], where "X" corresponds to the emerging device.

Major advantages of such an OIM as compared to other Ising machines like quantum annealing and optics-based coherent Ising machines (CIM) are room temperature operation, scalability (given OIM's CMOS compatibility), energy efficiency etc. Quantum annealers require extremely low operating temperatures, and quantum systems are also highly susceptible to errors due to decoherence[37,38]. CIMs can operate at room temperature, but they require the use of a long fibre ring cavity and a power-hungry field-programmable gate array (FPGA) to implement coupling thereby not making them scalable[25,37]. On the other hand, quantum annealers require extremely low operating temperatures, and quantum systems are also highly susceptible to errors due to decoherence[38,39]. It is to be noted here that the OIM has been deemed to be quantum-inspired (while still operating at room temperature) with binarized oscillator phases loosely replicating quantum superposition and oscillator-to-oscillator coupling loosely replicating quantum entanglement[21,22,40].

Given the aforementioned advantages of OIM (room temperature operation, scalability, energy efficiency, etc.), the most suitable application of a CMOS or CMOS + X OIM is an edge artificial intelligence (edge AI) or edge computing application[41], where the OIM, deployed at the edge device, solves a combinatorial optimization in real time. However, most optimization problems for which OIMs have been demonstrated as solvers thus far (MaxCut[2,11,28,42–44], Graph Colouring[45, 46], Maximum Independent Set[46,47], Traveling Salesman problem[30,46]) may not have such real time edge applications. Various other combinatorial optimization problems related to the domains of chemistry and biology (like drug discovery[48] and protein folding[49]), for which Ising machines have been considered in the past[12-20], do not demand real-time edge solutions. Hence, OIMs may not be needed for them (other IMs like quantum annealer or CIM can serve the purpose).

On the contrary, in this article, through a combination of device-level experiments and circuit-level simulations, we design a CMOS + X OIM for an actual edge application: solving the symbol detection problem in a multiple-input multiple-output (MIMO) wireless communication system. For the purpose, we have used CMOS circuits as ring oscillators and CMOS-compatible FeFET devices (fabricated using GlobalFoundries's 28 nm super low power (SLP) high-K/ metal gate (HKMG) CMOS technology[50,51]) as non-volatile memory (NVM) coupling elements between the oscillators (X).

MIMO technology, which involves multiple transmitter and receiver antennas (as shown in Fig. 1a), is extensively used in modern wireless communication systems[50, 51] because it offers the advantages of diversity gain (which mitigates information fading due to obstacles) and spatial multiplexing gain (which leads to enhanced data rate of communication)[52-54]. The symbols received in a MIMO system (for example $y_1$, $y_2$ in Fig. 1a for a 2 × 2 MIMO system) can be different from the original transmitted symbols ($x_1^t$, $x_2^t$ in Fig. 1a) due to the influence of the channel and associated noise. Determining the original transmitted symbols based on the received symbols is known as the MIMO symbol detection/ decoding problem[55–60]. Since this decoding needs to happen real time near the receiving antennas of the MIMO system, MIMO symbol decoding is an interesting edge computing application[43,61,62].

However, implementing MIMO symbol decoding on conventional classical digital hardware (CPUs, GPUs, other digital hardware for the edge) takes significant time. The most accurate way of MIMO symbol decoding (including low SNR regimes) is the exact/ brute-force maximum likelihood (ML) detection technique. This method works as follows: given a received vector **y** at the receiving antennas of the MIMO system, all possible transmitted vectors **x** are generated in a brute-force way, and among them, a particular vector is selected such that the square of the Euclidean distance (ED) between **y** and the product of the transmitted vector **x** and the channel coefficient matrix (**H**) (the product **Hx** corresponds to the vector that would have been received if there was no noise) is minimized (more details in Results Section). Thus, ML-based MIMO detection is a NP-Hard combinatorial optimization problem that optimizes this Euclidean distance (ML objective function), and the time taken to solve it on conventional classical digital computing hardware grows exponentially with a linear increase in MIMO system size (number of transmitting and receiving antennas) given the exhaustive nature of the brute-force ML detector[55,56]. Approximate ML detection solvers like sphere decoder (SD) and its different variations reduce the search space but still have time complexity varying between polynomial and exponential with respect to MIMO size[63–65]. Linear decoders implementable on conventional digital hardware, like zero-forcing (ZF) and minimum mean-squared error (MMSE) algorithms, also exist. But they also only offer polynomial time complexity given that they rely on matrix inversion for their operations[66].

In this article, we design a CMOS + X OIM that solves the ML decoding problem heuristically by obtaining the ground state or close-to-ground state of the classical Ising Hamiltonian[2,11] corresponding to the Euclidean distance (ED) term in the ML-based formulation of the MIMO symbol decoding problem (more details in later sections). We show that the computation time of the OIM (as defined by the time taken by the oscillator phases to settle such that their binarized values correspond to the final solution) grows only logarithmically (O($log(M)$)) with MIMO size $M$ ($M \times M$ MIMO considered here, size $M$ = number of transmitting antennas = number of receiving antennas). On the other hand, computation time of exact/ brute-force ML detector (implemented on conventional digital hardware) grows exponentially ($2^M$) and that of ZF, which offers one of the best time complexities among the approximate MIMO decoders discussed above, grows cubically with MIMO size (O($M^3$)). We show the logarithmic growth of computation time with MIMO size of the designed OIM through SPICE circuit simulations (with experimentally calibrated FeFET device characteristics embedded in them) and then further justify this logarithmic time complexity analytically using the Kuramoto model of coupled oscillators[2,11]. This is our major finding in this article since it demonstrates the suitability of the designed CMOS+X hardware for an edge computing application like MIMO symbol decoding.

We justify the choice of FeFET as the coupling element in the OIM by showing that the OIM's bit error rate (BER) increases significantly (implying significant performance drop) for any conductance modulation range of the coupling element outside 1 $\mu S$ to 100 $\mu S$ range and by also showing that only the FeFET device (as measured by us experimentally here) offers the above conductance modulation range among various emerging NVM devices studied recently, including spintronic devices and different kinds of resistive random-access memory (RRAM) devices[67–71]. This is another major finding in this article. Though FeFET has been demonstrated as the coupling element in OIMs in other reports[32,33], justification of its use because of its conductance modulation range has not been provided in these works though similar range of conductance modulation as we measure here has been reported before[31,72]. Also, these other articles on FeFET-based OIM[32,33] do not explore the MIMO symbol decoding problem (unlike us) and explore other optimization problems instead.

With respect to the MIMO symbol decoding problem, results for Kuramoto-model-based numerical simulation of an OIM on conventional classical digital computers (CPU and GPU) and its emulation on a custom-fabricated digital electronic chip have been reported very recently[73, 74]. But to the best of our knowledge, no design, simulation, or experimental demonstration of physical implementation of a MIMO-symbol-decoding OIM through analog electronic CMOS + X hardware has been published thus far. Also, no time complexity advantage of OIMs for the MIMO symbol decoding problem has been reported earlier to the best of our knowledge.

**Results**

**Overview of our OIM design and simulation:** In the MIMO symbol decoding problem (as shown in Fig. 1a), based on the received vector **y**, the transmitted vector **x** needs to be estimated. For this, based on **y** and our knowledge of the channel coefficient matrix **H**, we obtain the expressions for the linear and quadratic terms of the aforementioned square of ED term and thereby formulate the Ising Hamiltonian matrix (detailed derivation in the next sub-section).

Next, we design a coupled oscillator array circuit as the OIM (as shown in Fig. 1a) where CMOS-based blocks serve as the ring oscillators and all-to-all connectivity between the oscillators is achieved through a crossbar array of FeFET devices. Such all-to-all connected design suits the dense nature of the Ising Hamiltonian matrix obtained in the next sub-section. Programming of the FeFET coupling elements is carried out such that conductance values of these devices (for example, $G_{01}^+, G_{02}^+, G_{12}^+, G_{01}^-, G_{02}^-, G_{12}^-$ in the 2 × 2 MIMO system of Fig. 1a) are proportional to the coefficients of the aforementioned linear and quadratic terms in the expression for square of ED. A BSIM4 model of the conventional silicon transistor, calibrated against published data from the industry on 28 nm CMOS technology[50], is used to model the conventional transistors in the ring oscillators of our design. We choose FeFET as the NVM coupling element between any two oscillators and experimentally report, for that purpose, high linear range of conductance variation (1 μS to 100 μS approximately) upon the application of programming voltage pulses on a HfO$_2$-based FeFET device fabricated at Global Foundries's 28 nm high-K/ metal gate (HKMG) technology[50,51] (details in later sub-sections). We incorporate this experimentally measured conductance modulation characteristic in our SPICE simulation of the designed OIM circuit.

Performance of the designed OIM circuit, as obtained through this SPICE simulation, is presented in this article with the help of various metrics as summarized next (details in later sub-sections). Following the evolution of the phases of the oscillators over time, phases of the other oscillators relative to the reference oscillator ($\phi_1$, $\phi_2$ in Fig. 1a) are rounded off to 0 or $\pi$. Given that we design our OIM for Binary Phase Shift Keying (BPSK) modulation[52], if the phase is 0, the predicted transmitted BPSK symbol is +1. If the phase is $\pi$, the predicted transmitted BPSK symbol is -1. The predicted BPSK symbols are next compared with the actual transmitted BPSK symbols to obtain Euclidean distance, BER and success probability (SP)[52]. We show in a later sub-section that accuracy of the designed OIM, as gauged through the above three metrics, doesn't drop significantly up to MIMO size (*M*) of 100 for an optimum choice of design parameters. Also, we show that the computation time of the OIM grows only logarithmically with MIMO size $M$ as mentioned earlier.

**ML-based formulation of the MIMO symbol decoding problem for OIM design:** We model a MIMO system comprising $M$ transmitting and $M$ receiving antennas ($M \times M$) as[52,75]:

$$\mathbf{y} = \mathbf{H}\mathbf{x} + \mathbf{n}, \qquad (1)$$

where $\mathbf{y} \in \mathbb{C}^{M \times 1}$ denotes the received signal vector, $\mathbf{H} \in \mathbb{C}^{M \times M}$ represents the channel matrix that characterizes the fading effects between the transmitter and receiver, $\mathbf{x} \in \{+1, -1\}^{M \times 1}$ is the transmitted signal vector (since we consider BPSK modulation), and $\mathbf{n} \in \mathbb{C}^{M \times 1}$ corresponds to the additive white Gaussian noise (AWGN) vector. Since $\mathbf{H}$ and $\mathbf{y}$ are complex, they can be decomposed into real ($\mathbf{H_I}$, $\mathbf{y_I}$) and imaginary ($\mathbf{H_Q}$, $\mathbf{y_Q}$) components:

$$\mathbf{H} = \mathbf{H_I} + j\mathbf{H_Q}, \qquad \mathbf{y} = \mathbf{y_I} + j\mathbf{y_Q}. \qquad (2)$$

Each element of $\mathbf{H_I}$ ($H_I[i,j]$) and $\mathbf{H_Q}$ ($H_Q[i,j]$) represents the fading gain (including both amplitude and phase shift) of the channel between the $j$-th transmitting antenna and the $i$-th receiving antenna. Each such element is an independent and identically distributed random variable with the associated probability density function (PDF) corresponding to the Rayleigh fading channel:

$$P(h, \sigma) = \frac{h}{\sigma^2} e^{-\frac{h^2}{2\sigma^2}}, (h \geq 0), \qquad (3)$$

where $h = \sqrt{H_I[i,j]^2 + H_Q[i,j]^2}$, and $\sigma$ is the scale parameter[52,76,77].

The MIMO symbol decoding problem is about determining the transmitted vector **x** based on the received vector **y**. In the brute-force exact maximum likelihood (ML) detection method of MIMO symbol decoding, all possible transmitted vectors **x** in the $\{+1,-1\}^M$ space are generated one by one, multiplied with the channel matrix **H** (which is known), and then the square of the L2 norm of the difference between the received vector **y** and **Hx** ($f(\mathbf{x}) = (\|\mathbf{y} - \mathbf{Hx}\|_2)^2$), which is same as square of ED between **y** and **Hx**, is calculated. Whichever value of **x** yields the lowest $f(\mathbf{x})$ is determined to be the transmitted vector. Next, we derive the final expression for f(**x**) that needs to be optimized:

$f(\mathbf{x}) = (\|\mathbf{y} - \mathbf{H} \cdot \mathbf{x}\|_2)^2$

$= (\|(\mathbf{y_I} + j\mathbf{y_Q}) - (\mathbf{H_I} + j\mathbf{H_Q})\mathbf{x}\|_2)^2$

$= (\mathbf{y_I} - \mathbf{H_I x})^T (\mathbf{y_I} - \mathbf{H_I x}) + (\mathbf{y_Q} - \mathbf{H_Q x})^T (\mathbf{y_Q} - \mathbf{H_Q x})$

$= \mathbf{y_I}^T \mathbf{y_I} - 2\mathbf{y_I}^T \mathbf{H_I x} + \mathbf{x}^T \mathbf{H_I}^T \mathbf{H_I x} + \mathbf{y_Q}^T \mathbf{y_Q} - 2\mathbf{y_Q}^T \mathbf{H_Q x} + \mathbf{x}^T \mathbf{H_Q}^T \mathbf{H_Q x}$

$= \mathbf{y_I}^T \mathbf{y_I} + \mathbf{y_Q}^T \mathbf{y_Q} - 2(\mathbf{y_I}^T \mathbf{H_I} + \mathbf{y_Q}^T \mathbf{H_Q})\mathbf{x} + \mathbf{x}^T (\mathbf{H_I}^T \mathbf{H_I} + \mathbf{H_Q}^T \mathbf{H_Q})\mathbf{x}.$

Since the above function is to be optimized with respect to **x**, the first two terms in the function ($\mathbf{y_I}^T \mathbf{y_I}$ and $\mathbf{y_Q}^T \mathbf{y_Q}$), which only add a constant value to the function, can be ignored. Given that **x**, **y_I**, and **y_Q** are *M*-dimensional vectors and **H_I** and **H_Q** are $M \times M$ matrices, the remaining terms in $f(\mathbf{x})$ can be expressed in the summation form as follows (this is the ML objective function):

$$f(x) = -\sum_{i=1}^{M} a_i x[i] - \sum_{i=1, i<j}^{M} b_{ij} x[i] x[j], \qquad (4)$$

where **x**= $(x[1], x[2], ..., x[M])$ (where for all values of *i* from 1 to $M$, $x[i] \in \{-1, +1\}$). $a_i$ and $b_{ij}$ (coefficients of the linear and the cross terms respectively) are given as:

$$a_i = 2\sum_{k=1}^{M}(y_I[k]\boldsymbol{H_I}[k,i] + \boldsymbol{y_Q}[k]\boldsymbol{H_Q}[k,i]), \quad (5)$$

and

$$b_{ij} = -\sum_{k=1}^{M}(\boldsymbol{H_I}[k,i]\boldsymbol{H_I}[k,j] + \boldsymbol{H_Q}[k,i]\boldsymbol{H_Q}[k,j]). \quad (6)$$

Thus, minimization of $f(\mathbf{x})$ with respect to $\mathbf{x}$ is equivalent to finding out the spin configuration $\sigma = (\sigma_1, \sigma_2, ..., \sigma_M)$ (where for all values of $i$ from 1 to $M$, $\sigma_i \in \{-1, +1\}$) such that the following energy with respect to the classical Ising model for interacting spins (classical Ising Hamiltonian) is minimized:

$$E = \sum_{i=1}^{M} h_i \sigma_i - \sum_{i=1}^{M}\sum_{j=1(i<j)}^{M} J_{ij}\sigma_i\sigma_j, \quad (7)$$

where $h_i$ (proportional to $a_i$ as per equation 4 and equation 5) is the coefficient corresponding to the interaction between the spin and the external field (Zeeman energy term), and $J_{ij}$ (proportional to $b_{ij}$ as per equation 4 and 6) is the coefficient corresponding to the spin-spin interaction. An OIM meant to solve MIMO symbol decoding using ML detection will basically attempt to find the ground state of the Ising Hamiltonian in equation 7 by minimizing the Lyapunov function related to it through coupled oscillator dynamics[2,11,73,74]. We design this OIM next.

**Design of CMOS-based ring oscillators in the OIM:** Electronic ring oscillator circuits have been designed to implement an OIM in the past to solve the MaxCut problem[20–22,44] (but not the MIMO symbol decoding problem to the best of our knowledge). Ising Hamiltonian for the MaxCut problem only contains the quadratic term and not the linear term. But since the Ising Hamiltonian for the MIMO symbol decoding problem contains both the quadratic and the linear terms (equations 4 - 6), the ring oscillator IM design in previous reports[20–22,44] needs to be modified as we show next in this section of the paper.

An electronic ring oscillator circuit comprises an odd number of cascaded CMOS inverters, with the output of the last inverter connected to the input of the first inverter as feedback (Fig. 1b)[20–22,44]. For each such ring oscillator block in our Ising machine design, the input terminal of the first inverter is designated as the input terminal of the overall block, and the output terminal of the first inverter is designated as the output terminal of the block, as shown in Fig. 1a, b.

Using a SPICE circuit simulator[78,79], we design each ring oscillator block by connecting 27 CMOS inverters in a loop. The transistors of each inverter are based on the publicly available BSIM4 transistor model[80,81]. The natural frequency of each ring oscillator, as obtained from our SPICE simulation, turns out to be ≈ 1.24 GHz (the corresponding Fast Fourier Transform (FFT) spectrum of the time-dependent voltage waveform at the output terminal of the ring oscillator block (Fig. 1(b)) is shown in the Section S1 of Supplementary Information accompanying this article. The number of inverter stages in the ring oscillator block is chosen to be rather high (27) in order to limit the frequency of the oscillator to ≈ 1 GHz range because ring-oscillator-based OIM operating in this frequency regime has already been experimentally[21]. Operating frequency above this range may make the design of the circuit challenging for experimental implementation. More details of our CMOS-based ring oscillator circuit design can be found in the Methods section.

**Mapping of coefficients of ML objective function to coupling conductance values of OIM:** Next, we discuss the most important part of our design: how the coefficients $a_i$ and $b_{ij}$ of the ML objective function (square of ED) Ising Hamiltonian corresponding to the MIMO symbol decoding problem (equations 4 - 6) are mapped into our designed OIM. As shown in Fig. 1a, the ring oscillator blocks in our designed OIM are connected to each other in an all-to-all fashion through conductive coupling elements forming a crossbar array similar to that used in computing in memory or neuromorphic computing architectures[82-85]. The conductance values of the coupling elements are programmable and are proportional to the coefficient values: $a_i$ and $b_{ij}$. Such all-to-all connected design suits the dense nature of the Ising Hamiltonian matrix obtained for this MIMO symbol decoding problem (coefficients $a_i$ and $b_{ij}$ in equations 4 - 5 take non-zero values mostly as we will see in Fig. 2).

To implement a linear term with coefficient $a_i$ say, the $i$-th oscillator block is coupled to the reference oscillator block (indexed 0) through a pair of conductive elements with conductance value $G_{0i}$ ($G_{0i}$ proportional to $|a_i|$). To implement a quadratic term with coefficient $b_{ij}$ say, the $i$-th oscillator block is coupled to the $j$-th oscillator block through a pair of conductive elements with conductance value $G_{ij}$ ($G_{ij}$ proportional to $|b_{ij}|$).

Since coefficients $a_i$ and $b_{ij}$ can have both positive and negative values while conductance $G_{ij}$ can only be positive, we adopt the following strategy to implement coefficients of either sign. To implement coupling with positive coefficient values, the output terminal of the first oscillator block is coupled to the input terminal of the second block, and the input terminal of the first block is coupled to the output terminal of the second block[21,22]. If no other oscillator influences these two oscillators, then they will have a phase difference of $\pi$ in this case (out of phase). In Fig. 1a, the segment of the crossbar array labelled as "input-output cross bar" and having conductance values $G_{01}^+$, $G_{02}^+$ and $G_{12}^+$ takes care of this kind of coupling. To implement coupling with negative coefficient values, the output terminal of the first oscillator block is coupled to the output terminal of the second block, and the input terminal of the first block is coupled to the input terminal of the second block[21,22]. If no other oscillator influences these two oscillators, then they will have a phase difference of 0 in this case (in phase). In Fig. 1a, the segments of the crossbar array labelled as "input-input cross bar" and "output-output cross bar" and having conductance values $G_{01}^-$, $G_{02}^-$ and $G_{12}^-$ take care of this kind of coupling.

The mapping between the magnitude of the coefficient of any particular term (linear: $a_i$, quadratic: $b_{ij}$) (we call it $C$ say to generalize) and the conductance value $G$ of the corresponding coupling element ideally needs to be carried out based on the following equation:

$$G(C) = \frac{(C-C_{min})}{C_{max}-C_{min}}(G_{max}-G_{min}) + G_{min}, \qquad (8)$$

where $G_{min}$ and $G_{max}$ are the minimum and maximum conductance values exhibited by the coupling element, and $C_{min}$ and $C_{max}$ are the minimum and maximum coefficient magnitudes among all the coefficient magnitudes corresponding to all the linear terms ($a_i$, $i$ = 1 to $M$) and quadratic terms ($b_{ij}$, $i$ = 1 to $M$, $j$ = 1 to $M$) of the Ising Hamiltonian for a particular received symbol vector **y** in a given $M \times M$ MIMO system. It is to be noted that for a positive valued coefficient ($b_{ij}$), this mapping takes place between the coefficient magnitude $C$ and the conductance of the corresponding coupling element in the input-output crossbar array ($G^+_{i,j}$, $i < j$: Fig. 1(a)) while the conductance values of their counterparts in the input-input crossbar array and output-output crossbar array ($G^-_{i,j}$, $i < j$: Fig. 1(a)) are set to $G_{min}$ (to enable weakest coupling given that zero coupling isn't possible in this crossbar scheme). Similarly, for a negatively valued coefficient, mapping is carried out between $C$ and $G^-_{i,j}$ while $G^+_{i,j}$ is fixed at $G_{min}$. This mapping scheme has been summarized in Table 1.

The average values of $a_i$ (for all $i$ = 1 to $M$) and $b_{ij}$ ($i$ = 1 to $M$, $j$ = 1 to $M$) are plotted against MIMO size $M$ in Fig. 2 (a), (b) respectively. The variations of these values for a given MIMO size are plotted as red bars. The bottom of the bar corresponds to the minimum $a_i$ (or $b_{ij}$) value, and the top of the bar corresponds to the maximum $a_i$ (or $b_{ij}$) value, as shown. For a given MIMO size $M \times M$, the lower of the two values (minimum $a_i$ and minimum $b_{ij}$) is chosen as the minimum coefficient value $C_{min}$ in equation 8 above, and the higher of the two values (maximum $a_i$ and maximum $b_{ij}$) is chosen as the maximum coefficient value $C_{max}$ in equation 8 above.

The coupling elements of the crossbar array need to retain their conductance values until the next bit string is received and hence should preferably be made of non-volatile memory (NVM) devices. Also, using static random-access memory (SRAM) based coupling elements will not only lead to energy consumption (in order to retain the coupling values) but also increase the area footprint significantly (because of the high transistor count), which can be greatly reduced through the use of emerging NVM devices. But the conductance modulation characteristic of the chosen NVM device needs to follow certain conditions listed below (which we have identified and justified through this work as detailed later):

(i) The conductance range (difference between $G_{max}$ and $G_{min}$) should be as high as possible, preferably a few decades so that low magnitude of coupling coefficient corresponds to very weak coupling between the oscillators and high magnitude corresponds to strong coupling.

(ii) While equation 8 corresponds to linear mapping, NVM devices exhibit non-linearity in their conductance modulation (long-term potentiation (LTP) and long-term depression (LTD)) characteristics which will add to errors in the computation. So the linearity should be as high as possible, or rather the the magnitudes of the non-linearity factors $\alpha_P$ and $\alpha_D$ for LTP and LTD respectively, as defined in the manual for the Neuro Sim simulator related to NVM devices[86,87], should be as low as possible.

(iii) All conductance values between $G_{min}$ and $G_{max}$ should be in a suitable range so that the OIM operates desirably. Below these conductance values, coupling is too weak and the oscillators hardly interact. Above these conductance values, coupling is so strong that very high current from the second oscillator flows into the first oscillator and vice versa and the oscillations of both oscillators stop.

**Experimental measurement of conductance modulation in the FeFET device (coupling element between oscillators in OIM):** Next, we carry out experimental measurement of the conductance modulation in a $HfO_2$-based FeFET device[50,51]. The FeFET device used for this article has been fabricated based on GlobalFoundries's 28 nm SLP HKMG CMOS technology node as mentioned before. More details of the fabrication process can be found in the report by Trentzsch *et al*[50]. The schematic of the measured FeFET device is shown in Fig. 3a. It is a three-terminal device based on a metal (M)/ ferroelectric (F)/ insulator (I)/ semiconductor (S), or MFIS, structure as shown.

For conductance increase or LTP, at the metal gate of the device, we apply positive polarity programming or write pulses of increasing magnitude in this manner: pulse 1: 1 V, pulse 2: 1 V + $V_{step}$, pulse 3: 1 V + 2$V_{step}$, and so on, where $V_{step}$ = 109.4 mV (Fig. 3b). For conductance decrease or LTD, at the gate of the device, we apply negative polarity programming or write pulses of increasing magnitude in this manner: pulse 1: -1 V, pulse 2: -1 V + $V'_{step}$, pulse 3: 1 V + 2$V'_{step}$, and so on, where $V'_{step}$= -81.25 mV (Fig. 3c). The drain-to-source voltage ($V_{DS}$) is set at 0 V during programming.

After each LTP or LTD write pulse, a quasi-static voltage sweep (from -0.5 V to 1 V) is applied at the gate of the device as a read waveform, while fixing $V_{DS}$ at 100 mV. The measured drain current ($I_D$) at the end of different write pulses is plotted as a function of this applied gate-to-source voltage ($V_{GS}$) in Fig. 3d, f. For positive write pulses (LTP), the threshold voltage ($V_{th}$) gradually decreases with increasing magnitude of the pulse (Fig. 3d). Hence, for a given value of $V_{GS}$, $I_D$ and conductance $G$ (defined as $\frac{I_D}{V_{DS}}$) increases with increasing pulse number $N$: LTP. We plot such $G$ vs $N$ characteristics (LTP) for different $V_{GS}$ values in Fig. 3e. For negative write pulses (LTD), $V_{th}$ gradually increases with increasing magnitude of the pulse (Fig. 3f). Hence, in this case, for a given value of $V_{GS}$, $I_D$ and conductance G decreases with increasing pulse number: LTD. We plot such G vs N characteristics (LTD) for different $V_{GS}$ values in Fig. 3g.

Such negative threshold voltage shift ($V_{th}$ decreasing), leading to LTP, and positive $V_{th}$ shift, leading to LTD, have been reported experimentally in previous reports[31,72,88,89]. Such behaviour is observed experimentally due to the gradual switching of multiple ferroelectric domains, leading to shift of $V_{th}$ and resultant increase or decrease of conductance. Once all the domains have switched, conductance $G$ saturates leading to no further change with subsequent pulses (as observed after around 30 pulses in Fig.

3e). This experimentally observed phenomenon has also been modelled theoretically and through numerical simulations (Preisach's model of ferroelectric switching coupled with transistor models) in the above reports and other related reports[31,72,90-93].

For on-chip learning in crossbar arrays for computing in memory or neuromorphic computing applications, the conductance value of a similar NVM device, acting as a synapse, needs to be increased at some iterations and decreased at some other iterations to achieve training of the corresponding neural network in the crossbar array [31,83,84]. So, both LTP and LTD characteristics of the devices (preferably programmed through an identical pulse scheme in order to simplify the peripheral circuit design) need to be utilized. But in this OIM application, we need to program the coupling elements in the crossbar array of the designed OIM only once when a new **y** vector is received (Fig. 1a) as per equation 5 and 6 above. Hence, either LTP or LTD characteristic of the device can be used for programming all the coupling elements in the OIM. Also, using non-identical pulses (like we have used here: Fig. 3c) won't be quite sub-optimal in this case since programming is needed less frequently as explained above.

For all the LTP plots of Fig. 3e, the first few programming pulses do not change the conductance leading to high non-linearity of the LTP plots. Since the programming pulses used here for LTP (Fig. 3b) have gradually increasing amplitude, we observe the conductance to increase only after a certain number of pulses such that the pulse amplitude is finally larger than the minimum coercive voltage needed to switch the ferroelectric domains. The effect is found to be less pronounced in our reported LTD characteristics of Fig. 3g. One possible reason for this can be the difference in coercive voltage distribution of the domains with opposite ferroelectric polarizations. A broader distribution leads to a gradual or linear change as we observe for our LTD plots here[93].

Another reason can be the fundamental asymmetry of the gate stack for the FeFET device: MFIS structure. Hence, we don't use the LTP plots given their high non-linearity (criterion (ii) above violated).

Among the LTD plots of Fig. 3g, we don't use the plot for $V_{GS}$ = 0.5 V, 0.9 V, and 0.95 V because of the range of conductance modulation is not optimum (criterion (iii) above not satisfied suitably: more details in the next section and Table 2). Among the LTD plots for $V_{GS}$ = 0.6 V, 0.7 V, and 0.8 V, which all satisfy criterion (iii) (next section, Table 2), we choose the one for 0.6 V for every coupling element in the designed OIM circuit because it exhibits the highest linearity among the three (criterion (ii) above) and use it (after applying a linear fit: red dashed line in Fig. 3g) in the coupling element model of our designed OIM circuit (Fig. 1a) in SPICE, as presented in the next sub-section.

**Simulation results for CMOS + FeFET OIM (accuracy and computation time):** For different MIMO sizes $M$ ($M$ transmitting × $M$ receiving antennas), OIM circuits with ($M$+1) ring oscillators, as designed above (Fig. 1a) are simulated in SPICE (simulator details mentioned in Methods section). For each MIMO size $M$, 10-20 use cases are considered (Fig. 4). For each of the above use cases, $M \times M$ channel matrix coefficients $H_I[i,j]$ and $M \times M$ coefficients $H_Q[i,j]$ are generated such that $h$ follows the Rayleigh fading channel distribution (equation 3) with $\sigma$ = 1. For each use case, a different transmitted vector **x** in the $\{+1, -1\}^M$ space is considered, and using equation 1, the received vector **y** is calculated. Then, using equation 5 and 6, coefficients of the Ising Hamiltonian $a_i$ and $b_{ij}$ are calculated. Then these coefficients are mapped into conductance values of coupling elements in the crossbar array (Fig. 1a) in our SPICE design using the conductance characteristic shown through the dotted red line in Fig. 3g and following the mapping scheme in Table 1. Then for each use case of the OIM for *M* × *M* MIMO, SPICE simulation is carried out with the OIM circuit's dynamics evolving up to 2 $\mu$s with smallest time step for the simulation fixed at 100 ps.

After each simulation, we obtain time-dependent voltage signals at the output terminals of all the (*M* +1) ring oscillators. Using the voltage waveform of the reference oscillator as the reference, relative phases of all other oscillators are obtained and then rounded off (binarized) to even multiples of $\pi$ (same as 0) or odd multiples of $\pi$ (same as $\pi$). Given that phase 0 corresponds to BPSK symbol +1 and phase $\pi$ to BPSK symbol -1, the transmitted vector is estimated as $\mathbf{x}^{OIM}$ based on the phase values at the end of the simulation (or after phases become stable). One such example is shown in Section S2 of Supplementary Information where the symbol decoding problem is solved successfully using this method for one use case on the 3 ×3 MIMO system.

The following four metrics are calculated across all use cases for each considered MIMO size and plotted in Fig. 4:

(i) **Euclidean distance:** For a particular MIMO size *M*, with respect to each use case, SPICE simulation is performed, and the square of Euclidean distance between the estimated transmitted vector $\mathbf{x}^{OIM}$ and the actual transmitted vector $\mathbf{x}^{T}$ is calculated as $(\|\mathbf{x}^{OIM} - \mathbf{x}^{T}\|_2)^2$. The average Euclidean distance across all use cases is plotted as a function of MIMO size *M* in Fig. 4a. The standard deviation in Euclidean distance across all use cases is plotted as an error bar. No error bar for a particular MIMO size means standard deviation is 0 for the use cases corresponding to that size.

(ii) **Bit error rate (BER):** For a particular MIMO size *M*, with respect to each use case, SPICE simulation is performed, and BER is calculated as the ratio of the number of BPSK symbols in $\mathbf{x}^{OIM}$ that do not match with that in $\mathbf{x}^{T}$ to the total number of BPSK symbols in either vector ($M$). Average BER across all use cases is plotted as a function of MIMO size $M$ in Fig. 4b, and the standard deviation (when it is not 0) is shown as an error bar.

(iii) **Success Probability (SP):** For a particular MIMO size $M$, the ratio of number of use cases for which $\mathbf{x}^{OIM}$ matches exactly with $\mathbf{x}$ to the total number of use cases is the success probability (SP). SP is plotted as a function of MIMO size $M$ in Fig. 4c.

(iv) **Computation Time:** Time taken by the phases of the oscillators to reach values binarized versions of which (0 or $\pi$) remain unchanged since then till the end of simulation time (2 $\mu$s) and thereby yield the final solution is reported here as computation time. Following this definition, computation time corresponds to the minimum time the designed OIM needs to reach its final solution. Average computation time across total number of use cases is plotted vs MIMO size in Fig. 4d, while the standard deviation is plotted as an error bar like before. Given that the ring oscillators oscillate at ≈ 1.24 GHz, computation time can be divided by the oscillation time period (≈ 0.806 ns) to obtain the number of oscillation cycles it takes for the phases to reach steady state values.

From Fig. 4 a, b, c, we observe that the accuracy of the OIM (as measured in terms of Euclidean distance, BER, SP) remains almost unchanged from 2×2 MIMO up to 90 × 90 MIMO, with some performance drop observed for 100 × 100 MIMO. For all these results in Fig. 4, conductance range of the coupling element is taken to be 1 $\mu$S to 60 $\mu$S (as exhibited by the FeFET device: red plot in Fig. 3g) as mentioned before. In the Discussion section (and Table 2 and Table 3 associated with it), we analyse how accuracy of the OIM is affected for coupling conductance values outside this range and thereby justify the use of the FeFET device in our design.

Here, we also compare the growth of computation time of the designed OIM circuit with MIMO size, as plotted in Fig. 4d (corresponding to the experimentally observed 1 $\mu S$ to 60 $\mu S$ range of the coupling conductance), with exact/ brute-force ML detection algorithm and the zero forcing (ZF) algorithm both implemented on CPU. Among the various conventional classical approximation algorithms for MIMO decoding introduced earlier in this article, we have chosen ZF and numerically implemented it here because ZF offers one of the best time complexities among these algorithms. So, if we can show that time complexity for the designed OIM is better than ZF, then it implies OIM offers a faster alternative to other conventional algorithms as well for implementing MIMO symbol decoding.

For accurate comparison, computation time obtained for both ZF numerical implementation on CPU and CMOS + FeFET OIM circuit (as obtained from SPICE simulation) are divided by the computation time for these methods for 2 × 2 MIMO (as listed in Table 4). These normalized values are plotted versus MIMO size ($M$) in Fig. 5a. Computation time for exact/ brute-force ML has also been plotted based on theoretical calculation. Fig. 5a shows that computation time for the designed CMOS + FeFET OIM grows much slowly compared to that of ZF (polynomial growth for ZF[65]) and brute-force ML (exponential growth). The difference is more pronounced when we plot normalized computation time as a function of logarithm of MIMO size ($log_2(M)$) in Fig. 5b for all the methods. Only for the designed CMOS + FeFET OIM, computation time vs $log_2(M)$ for OIM shows linear behaviour (for the other methods, it shows exponential growth with $log_2(M)$). Thus, our SPICE simulation predicts that if the designed CMOS + FeFET OIM is implemented on actual hardware, then its computation time will grow only logarithmically with MIMO size while that for other methods grows either polynomially or exponentially. This makes the designed OIM a much faster alternative to conventional classical digital hardware to solve the MIMO decoding problem.

It is to be noted here that another related metric, time to solution (TTS), is often used to describe the speed of a wide range of Ising machines: Fujitsu digital annealer, Toshiba bifurcation machine, DWave quantum annealer, optical coherent Ising machine (CIM), etc.[10]. To calculate this TTS metric as per the definition in the report by Mohseni[10], the solutions are only restricted to the correct solution (or within a fixed error percentage with respect to the correct solution). But following this definition, TTS varies exponentially or near exponentially with problem size ($M$, e.g., size of the MIMO system in our case) for all Ising machines with distinction between the Ising machines lying only in the pre-factor $c$ as defined below: $TTS \propto e^{cM}$, or $TTS \propto e^{c\sqrt{M}}$. On the other hand, our definition of computation time here doesn't restrict the solution to the correct solution. Given that we focus on an edge computing problem here (MIMO symbol decoding needs to be performed real time), the time by which a solution is reached needs to be minimized here as much as possible. The solution quality can be interpreted from accuracy metrics like Euclidean distance, BER, and SP (Fig. 4) without the solution time calculation not relying on these metrics. Our definition of computation time here follows this criterion, and so we use this definition in our article. However, for other problems which need not be solved at the edge (mentioned in the Introduction section), as dealt with by the other aforementioned Ising machines[10], the definition of TTS in the report by Mohseni *et al*[10] may be more suitable. We justify the obtained logarithmic growth of computation time of the designed CMOS + FeFET OIM using analytical treatment of the coupled oscillators through Kuramoto model in the Discussion section.

**Discussion**

**Impact of conductance range of the coupling element on OIM's accuracy and justification for using FeFET device as a coupling element in the designed OIM:** For all the results in Fig. 4, conductance range of the coupling element is taken to be 1 $\mu$S to 60 $\mu$S (as exhibited by the FeFET device: red plot in Fig. 3g) as mentioned before. In Table 2, we show how BER changes if maximum conductance is still two orders higher than minimum conductance like above but the absolute values of the conductance are in a different range. For a given MIMO size, the same 10 use cases are considered as before (Fig. 4). But now, the SPICE simulation of OIM is carried out for different conductance ranges of the coupling element: 1 nS to 100 nS, 10 nS to 1 $\mu$S, 100 nS to 10 $\mu$S, 1 $\mu$S to 100 $\mu$S, 10 $\mu$S to 1 mS, 100 $\mu$S to 10 mS, and 1 mS to 100 mS. From Table 2, we observe that for small MIMO sizes, different coupling conductance ranges can yield reasonably low BER. But for larger MIMO sizes, only 1 $\mu$S to 100 $\mu$S yields low BER. If the coupling conductance takes values lower than this range, the oscillators are very weakly coupled leading to high BER. If the coupling conductance takes values significantly higher than this range (e.g., 100 $\mu$S to 10 mS, and 1 mS to 100 mS), the net current flowing into one ring oscillator block from other ring oscillator blocks is so high that the oscillations cease to exist and BER can't be reported. So, the circuit can no longer be used as OIM.

Table 3 shows the conductance ranges experimentally reported in various emerging NVM devices mentioned in Introduction section[31,67-72]. We observe that

*FeFET-coupled CMOS Ring Oscillator Array for MIMO Symbol Detection* 26among these devices, only the FeFET device, as reported here and in previous articles by Jerry *et al.*[31,72], offers the suitable conductance range for obtaining low BER (as identified above and in Table 2: 1 $\mu$S to 100 $\mu$S approximately) among these devices. This justifies the use of FeFET (as opposed to other emerging NVM devices) in our OIM design.

**Justifying logarithmic time complexity of designed OIM using Kuramoto model of oscillators:** Next, we explain the logarithmic growth of computation time (logarithmic time complexity) of OIM with MIMO size analytically using the Kuramoto model of oscillators: a physics-agnostic generalized model that has been applied in the past for a wide range of oscillators (electrical, mechanical, chemical, etc.)[2,11]. In order to analytically explain the above SPICE simulation results of the designed ring oscillator circuit very accurately, the circuit should be modelled using transistor device equations and Kirchoff's current law (KCL) and Kirchoff's voltage law (KVL). However, past analysis using such equations has shown that the behaviour of the coupled ring oscillator circuit (and other electronic oscillator circuits) is consistent with Kuramoto model[94-96].

To solve symbol decoding in MIMO system of *M* transmitters and *M* receivers, one reference oscillator and *M* other oscillators are needed (as discussed earlier). All oscillators are considered to have the same natural frequency. According to the Kuramoto model, the phases of the *M* oscillators ( $[\phi_1(t),\phi_2(t),...,\phi_M(t)]^T \in \mathbb{R}^M$) evolve over time as:

$$\frac{d\phi_i}{dt} = -\sum_{j=1,j\neq i}^{M} K_{i,j} \sin\left(\phi_i(t) - \phi_j(t)\right) - K_{i,ref} \sin\left(\phi_i(t) - \phi_{ref}\right), \qquad (9)$$

where $i = 1, 2, …M$. $\phi_{ref}$ refers to the fixed phase of the reference oscillator. $K_{i,j}$ represents the coupling strength between two oscillators indexed $i$ and $j$, and $K_{i,ref}$ represents the coupling between $i$-th oscillator and the reference oscillator.

The fixed points for this non-linear dynamic system are: $\phi^*_i(t) = \phi_{ref} + n_i\pi$, where $n_i$ can be an odd or even integer ($n_i \in \mathbb{Z}$). So, $\phi_i(t)$ can be expressed as $\phi^*_i(t) + \delta\phi_i(t)$, where $\delta\phi_i(t)$ corresponds to the deviation from the fixed point. Thus, equation 9 can be instead expressed as:

$$\frac{d(\delta\phi_i(t))}{dt} = -\sum_{j=1, j\neq i}^{M} K_{i,j} \sin\left((n_i - n_j)\pi + \delta\phi_i(t) - \delta\phi_j(t)\right) - K_{i,ref} \sin(n_i\pi + \delta\phi_i(t)). \quad (10)$$

Assuming small deviations from a fixed point and hence locally linearizing equation 10 around the fixed point using the Jacobian matrix technique[97-99], we obtain:

$$\begin{bmatrix} \frac{d\delta\phi_1(t)}{dt} \\ \frac{d\delta\phi_2(t)}{dt} \\ \vdots \\ \frac{d\delta\phi_M(t)}{dt} \end{bmatrix} = \begin{bmatrix} \{-\sum_{j=2}^{M} K_{1j} \cos\left((n_1-n_j)\pi + \delta\phi_1(t) - \delta\phi_j(t)\right) - K_{1,ref}\cos(n_1\pi + \delta\phi_1(t))\} & K_{1,2}\cos\left((n_1-n_2)\pi + \delta\phi_1(t) - \delta\phi_2(t)\right) & \cdots \\ K_{2,1}\cos\left((n_2-n_1)\pi + \delta\phi_2(t) - \delta\phi_1(t)\right) & \{-\sum_{\substack{j=1 \\ j\neq 2}}^{M} K_{2j} \cos\left((n_2-n_j)\pi + \delta\phi_2(t) - \delta\phi_j(t)\right) - K_{2,ref}\cos(n_2\pi + \delta\phi_2(t))\} & \cdots \\ \vdots & \vdots & \ddots \end{bmatrix} \begin{bmatrix} \delta\phi_1(t) \\ \delta\phi_2(t) \\ \vdots \\ \delta\phi_M(t) \end{bmatrix} = J \begin{bmatrix} \delta\phi_1(t) \\ \delta\phi_2(t) \\ \vdots \\ \delta\phi_M(t) \end{bmatrix},$$

(11)

where $J$ is the Jacobian matrix for that fixed point.

Given that $J$ is a real symmetric matrix, all its eigen values are real. For the fixed point to be a final steady-state solution of the system, it has to be an attractor state (provided this local linearization assumption), which means all the eigen values of $J$ have to be real and negative[97-99]. Hence, the general solution for equation 11 about this attractor fixed point is[99]:

$$\begin{bmatrix} \delta\phi_1(t) \\ \delta\phi_2(t) \\ \vdots \\ \delta\phi_M(t) \end{bmatrix} = \sum_{k=1}^{M} c_k \mathbf{v}_k e^{-|\lambda_k|t}, \qquad (12)$$

where $-|\lambda_k|$ corresponds to each eigen value of $J$ (real and negative, hence expressed as $-|\lambda_k|$) and $\mathbf{v}_k$ corresponds to the eigenvector of $J$ for that eigen value. $c_k$ is the coefficient for that eigen vector. $c_k$ is time-independent and depends only on the initial condition.

We next assume that the lowest $|\lambda_k|$ is significantly lower than the magnitudes of the other eigen values of the matrix. Let $k_{min}$ be the index for this eigen value. Hence, the term corresponding to $k_{min}$ (slowest decay) in the summation in the right-hand side of equation 12 dominates the overall sum. Hence, for large $t$ (significant passage of time), we can approximate:

$$\begin{bmatrix} \delta\phi_1(t) \\ \delta\phi_2(t) \\ \vdots \\ \delta\phi_M(t) \end{bmatrix} \approx c_{k_{min}} \mathbf{v}_{k_{min}} e^{-|\lambda_{k_{min}}|t}. \qquad (13)$$

The time $t_{comp}$ by which the system evolves close enough to the fixed point can be considered to be the computation time of the OIM because then the binarized phases of the oscillators will converge to that in the final solution. Thus, for $t \geq t_{comp}$,

$$(||\delta\phi(t)||_2)^2 \leq \epsilon, \qquad (14)$$

where $\epsilon \in R^+$.

From equation 13,

$$(||\delta\phi(t)||_2)^2 = c_{k_{min}}^2 (||\mathbf{v}_{k_{min}}||_2)^2 e^{-2|\lambda_{k_{min}}|t}. \qquad (15)$$

Using equation 13 and 15,

$$e^{2|\lambda_{k_{min}}|t} \geq \frac{c_{k_{min}}^2(\|\mathbf{v}_{k_{min}}\|_2)^2}{\epsilon} \tag{16}$$

Hence,

$$t \geq \frac{1}{2\lambda_{k_{min}}} \ln \left(\frac{c_{k_{min}}^2(\|\mathbf{v}_{k_{min}}\|_2)^2}{\epsilon}\right) \tag{17}$$

Since $(\|\mathbf{v}_{k_{min}}\|_2)^2$ is the sum of squares of *M* variables, after assuming that each such variable is a Gaussian random variable, $(\|\mathbf{v}_{k_{min}}\|_2)^2$ is a random variable that follows the Chi-squared distribution[100,101]. If *M* is large, expectation value of $(\|\mathbf{v}_{k_{min}}\|_2)^2$ is equal to *M*[100,101]. Hence,

$$t \geq \frac{1}{2\lambda_{k_{min}}} \ln \left(\frac{c_{k_{min}}^2 M}{\epsilon}\right) \tag{18}$$

Based on equation 18 and the aforementioned definition of computation time of OIM ($t_{comp}$), we can conclude $t_{comp}$ to be O(log(*M*)) (logarithmic time complexity). This explains the logarithmic growth of computation time with MIMO system size *M* as obtained from SPICE circuit simulations of the OIM in Fig. 5.

**Conclusion**

Thus, in this article, through a combination of device-level experiments and circuit-level simulations, we have designed a CMOS + X (FeFET) OIM specifically to solve an optimization problem at the edge: symbol decoding in MIMO wireless communication. Given that such CMOS + X hardware is scalable, operational at room temperature, and energy-efficient, it is more suitable for such edge computing applications compared to various other kinds of Ising machines. Among various emerging NVM devices (X), we have identified the FeFET device to be the most appropriate coupling element for this application since the conductance range it offers leads to lowest BER of the OIM. We have experimentally demonstrated that particular conductance range on a FeFET device fabricated at GlobalFoundries's 28 nm SLP HKMG technology node. Additionally, we have shown through SPICE circuit simulations and analytical treatment (using the Kuramoto model of oscillators) that computation time of the designed OIM scales only logarithmically with MIMO system size. On the other hand, solving the MIMO decoding problem exactly through a brute-force approach on a conventional classical digital computer (CPU or GPU) involves exponential time complexity and solving the problem approximately on the same hardware involves polynomial time complexity at best. Thus, our designed OIM offers much faster solution than conventional digital hardware making it highly suitable for real time problem solving at the edge like the particular application we consider here: MIMO symbol decoding.

**Methods**

**OIM Circuit Design and Simulation:** We have used the LTSpice SPICE circuit simulator provided by Analog Devices[78, 79] for all the circuit design work in this article including individual ring oscillator block designs (Fig. 1b) and overall MIMO circuit designs (Fig. 1a). We design each ring oscillator block by connecting 27 CMOS inverters in a loop, the transistors of each inverter being based on the publicly available BSIM4 transistor model[80, 81]. The width of each pMOS transistor is chosen to be 1 $\mu$m and that of each nMOS to be 600 nm. The length of both pMOS and nMOS transistors is chosen to be 30 nm. Supply voltage $V_{DD}$ in each ring oscillator block is set at 1.2 V.

Before using this transistor model in the ring oscillator block, we calibrate the characteristic of individual nMOS transistor and that of the individual pMOS transistor against the experimentally measured characteristics for Global Foundries's 28 nm gate-first SLP HKMG CMOS technology, as reported by Trentzsch *et al*[50]. For this purpose, the transfer characteristic of the transistor (drain current $I_D$ vs gate-to-source voltage $V_{GS}$ at a given drain-to-source voltage $V_{DS}$) has been obtained from SPICE simulation of the transistor using BSIM4 model for different $V_{DS}$ values. On current ($I_{on}$) vs off current ($I_{off}$) characteristic has been obtained from that and calibrated against experimentally obtained plots of the same as shown in the report by Trentzsch *et al*[50].

Different use cases are considered for the overall OIM circuit (corresponding to different transmitted vectors and channel matrices in the MIMO system, as explained in the Results section), and accordingly conductance values of the coupling elements between the ring oscillators are chosen calibrated against experimental data measured from the FeFET device. For each use case, SPICE simulation is carried out with the OIM circuit's dynamics evolving up to 2 $\mu$s with smallest time step for the simulation fixed at 100 ps.

**Measurement on the FeFET device:** Electrical characterization of the device, including measurement of transfer characteristics and application of voltage pulses, is carried out using a semiconductor parameter analyser (Keysight B1500) and its associated waveform generation/ fast measurement unit (WGFMU) for programming. The programming pulse scheme and the method to measure conductance of the FeFET device in response to those pulses are described in detail in the Results section.

**Supplementary Information**

Supplementary Information document accompanying this main text contains our SPICE simulation results for an individual ring oscillator block and for an OIM circuit corresponding to 3 x 3 MIMO for a particular transmission use case.

**Acknowledgments**

D. B. acknowledges funding support from Science and Engineering Research Board, India (through project MTR/2023/000851), Ministry of Education, India (through project MoE-STARS/STARS–2/2023-0257), and Industrial Research and Consultancy Centre (IRCC), Indian Institute of Technology Bombay (IITB) (through faculty seed grant). A. G. and D. B. acknowledge research funding support from Fractal Analytics (Level 7, Commerz II International Business Park, Oberoi Garden City, Goregaon, Mumbai, Maharashtra 400063). H. M., S. D., and S. B. acknowledge funding support from the European Union within "Next Generation EU", the Federal Ministry for Economic Affairs and Climate Action (BMWK) on the basis of a decision by the German Bundestag and the State of Saxony with tax revenues based on the budget approved by the members of the Saxon State Parliament in the framework of "Important Project of Common European Interest - Microelectronics and Communication Technologies", under the project name "EUROFOUNDRY".

**Author contributions**

A. G., S. N., S. A., and D. B. developed the mathematical formalism. H. K. J, S, M, and D. B. carried out the circuit design and simulations. H. M., S. D., and S. B. fabricated the devices.

**Figures and Captions**

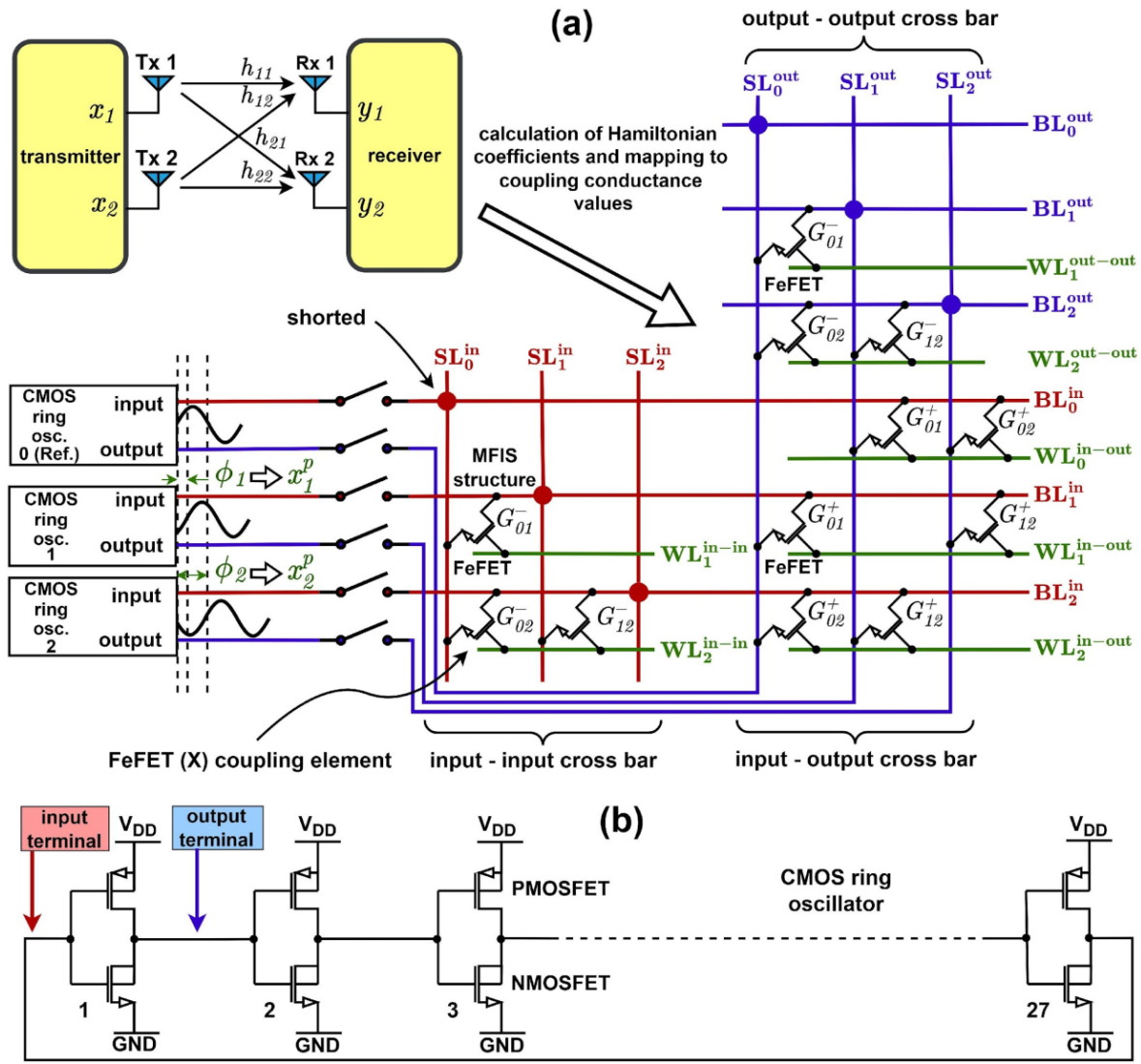

**Fig. 1. Schematics of our designed CMOS + FeFET OIM. a.** A schematic of a 2×2 MIMO wireless communication system and that of the OIM, meant to be implemented at the receiver for symbol decoding of the given MIMO system, are shown. The three ring

oscillators (a reference oscillator indexed 0 and two main oscillators indexed 1 and 2 corresponding to the two transmitters/ receivers of the MIMO system) are coupled through three-terminal FeFET devices with adjustable conductance values: $G_{01}^+, G_{02}^+$ and $G_{12}^+$ for positive coefficients in the corresponding Ising Hamiltonian and $G_{01}^-, G_{02}^-$ and $G_{12}^-$ for negative coefficients (more details in the text). These conductance values correspond to the conductance between source terminals of the FeFET devices and their drain terminals connected to source lines (SL) and bit lines (BL) of the crossbar array and are modulated by voltages applied at the gate terminals of the devices connected to the write lines (WL). The final phases of ring oscillator 1 and 2 with respect to the reference oscillator 0 are binarized to predict transmitted BPSK symbols $x^{OIM}[1]$ and $x^{OIM}[2]$ (phase 0 means BPSK symbol +1, phase $\pi$ means BPSK symbol -1). The predicted BPSK symbols are then compared with the actual transmitted BPSK symbols $x^T[1]$ and $x^T[2]$ to calculate BER and SP. **b.** Design of the CMOS-based ring oscillator circuit acting as an individual oscillator block in the OIM.

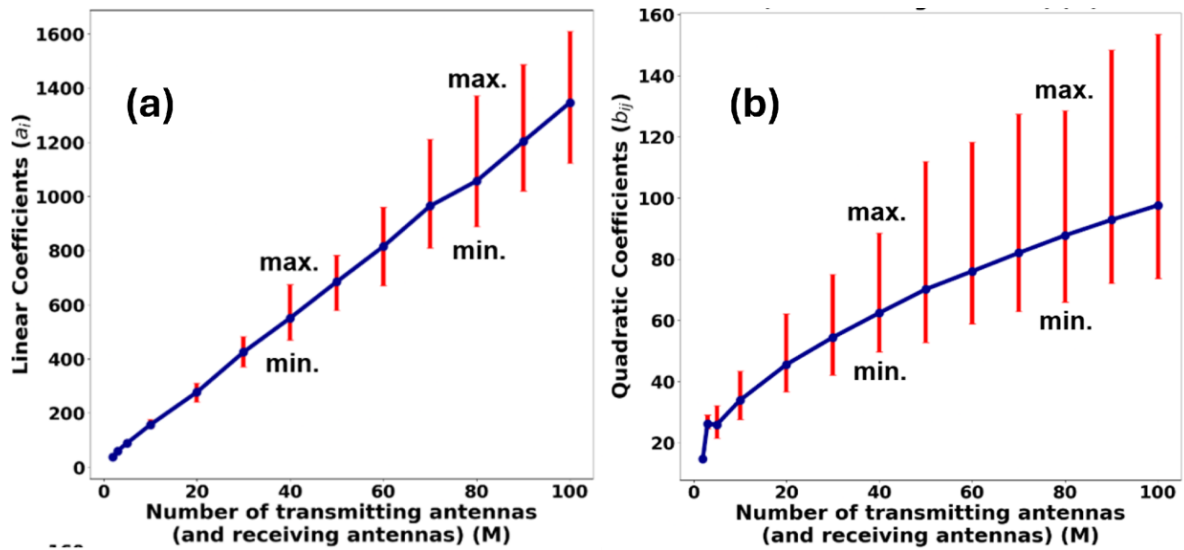

**Fig. 2. Values taken by linear coefficients ($a_i$) and quadratic coefficients ($b_{ij}$) in the objective function to be optimized under the ML formulation of MIMO decoding. a.** Average value of $a_i$ for a given MIMO size (average of $a_i$ values across several use cases corresponding to different transmitted vector values and different channel matrices) is plotted as a function of MIMO size *M*. **b.** Similar average value of $b_{ij}$ is plotted as a function of *M*. The variation of these values for each MIMO size across the different use cases is shown as a red bar.

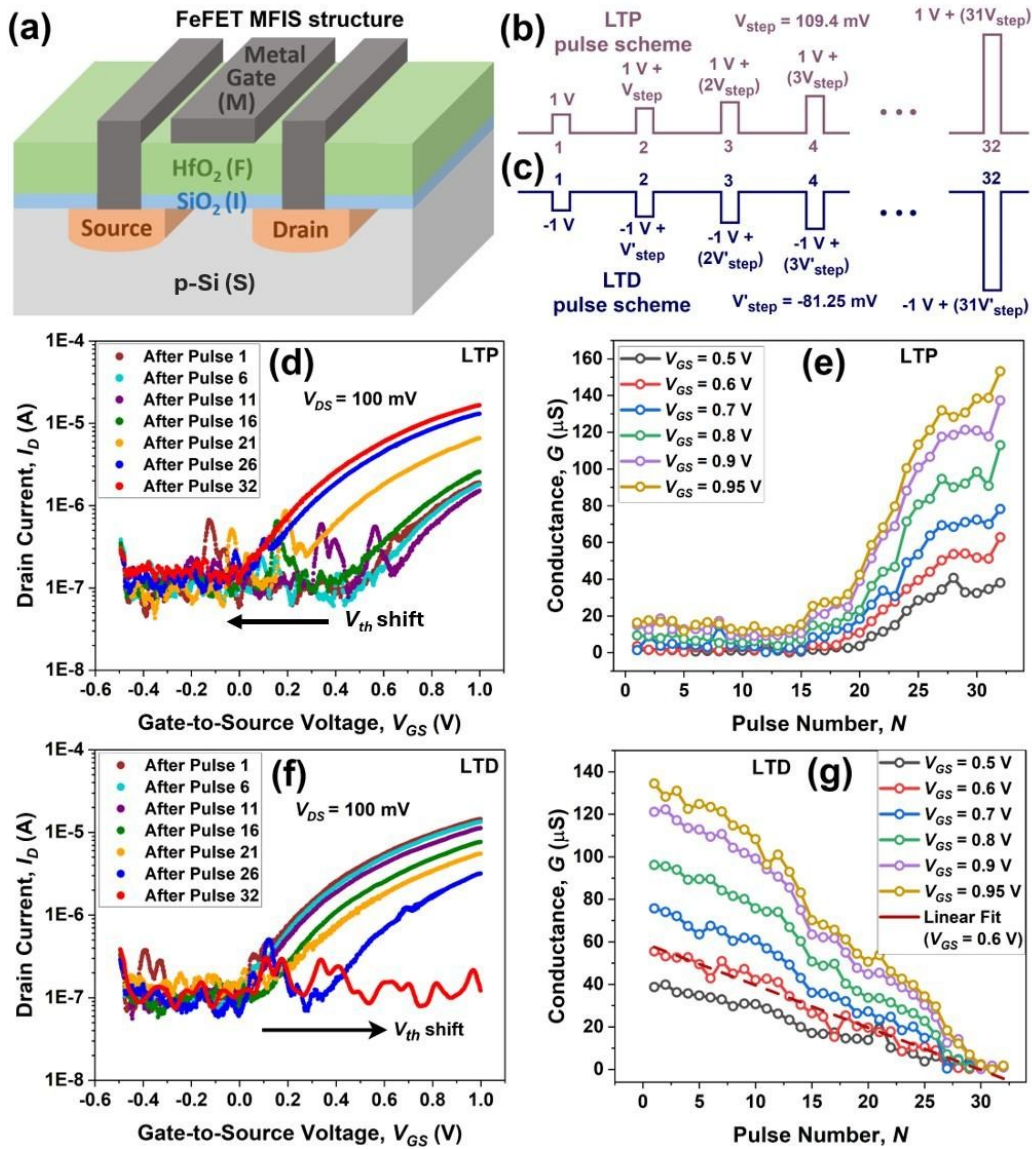

**Fig. 3. Experimental measurement of conductance modulation in the FeFET device (coupling element between oscillators in OIM): a.** Schematic of the fabricated FeFET device (fabricated at GlobalFoundries's 28 nm SLP HKMG CMOS technology node) following the metal-ferroelectric-insulator-semiconductor (MFIS) structure, with the ferroelectric being doped $HfO_2$, insulator being interfacial $SiO_2$, and semiconductor being p-doped silicon (Si). **b, c.** Programming pulse sequence at the gate to carry out gradual increase or decrease of conductance $G$ between source and drain: positive polarity pulses for LTP (b) and negative polarity pulses for LTD (c). **d, f.** Drain current $I_D$ vs gate-to-source voltage $V_{GS}$ during the read operation (with drain-to-source voltage $V_{DS}$ fixed at 100 mV) after applying a sequence of programming/ write pulses at the gate. **e, g.** Corresponding to the plots in (d), (f), ratio of $I_D$ to the fixed $V_{DS}$ is plotted here (for different values of $V_{GS}$) as drain-to-source conductance $G$ as a function of the program pulse number. (b), (d), (e) correspond to LTP, and (c), (f), (g) correspond to LTD.

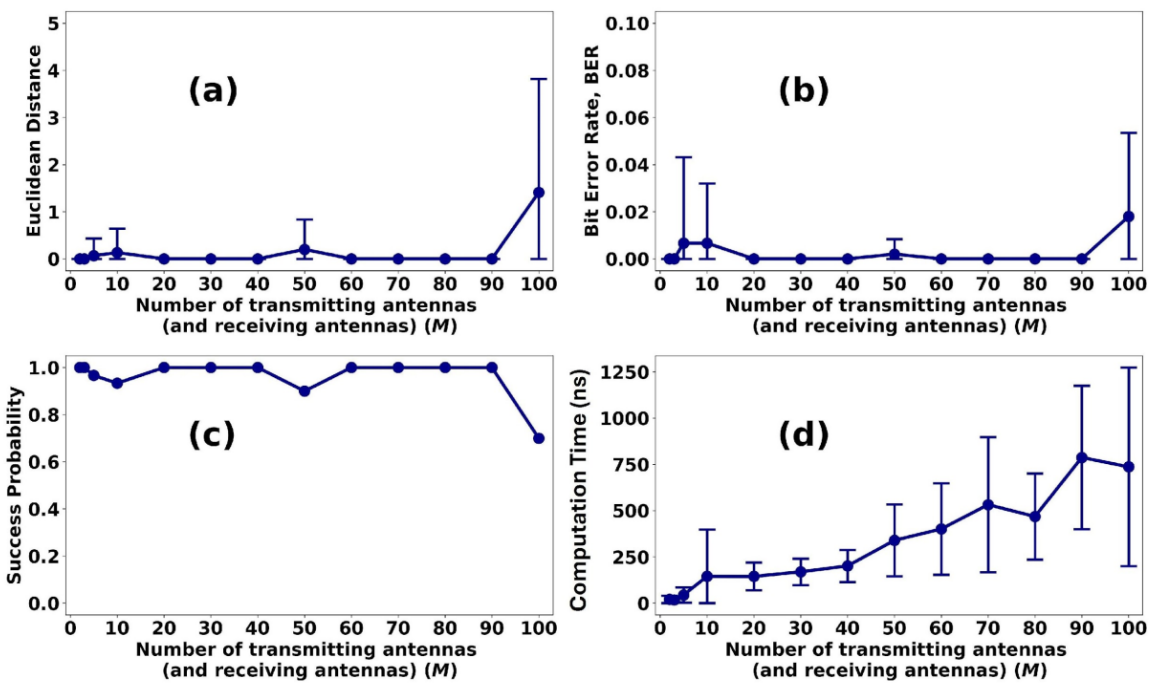

**Fig. 4. Performance-related results from SPICE simulation of the designed CMOS + FeFET OIM circuit for different MIMO sizes *M*.** With 10 use cases (for 10 dB SNR) (the meaning of an use case provided in the text) for *M* = 10, 20, 30, 40, and 50, and 20 use cases (for 10 dB SNR) for *M* = 60, 70, 80, 90, 100, average Euclidean distance is plotted in **(a),** average bit error rate (BER) is plotted in **(b),** success probability (SP) is plotted in **(c),** and computation time is plotted in **(d).** All these metrics are defined in the text. The standard deviation of each metric across the considered use cases is shown through an error bar (no error bar means standard deviation 0).

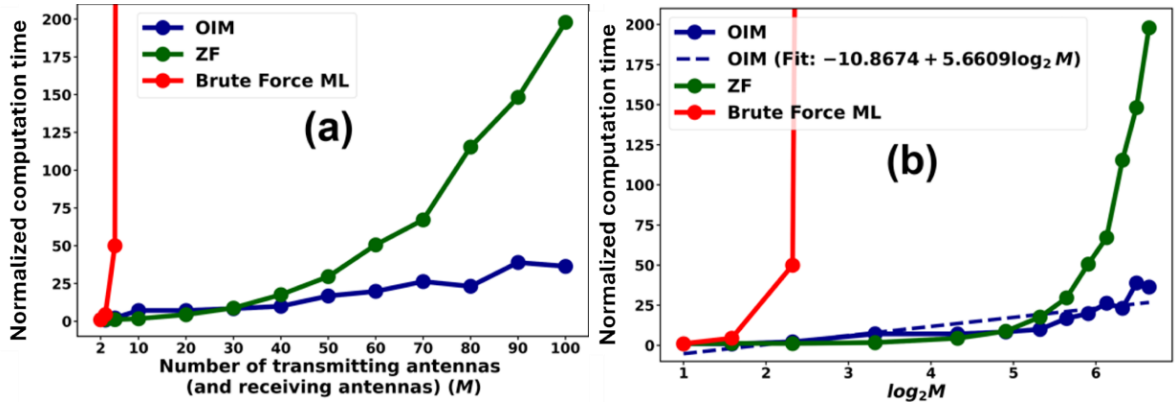

**Fig. 5. Logarithmic time complexity of the designed OIM**. **a.** Normalized computation time versus MIMO system size (*M*) for conventional classical methods implemented on CPU (exact/ brute-force ML, ZF) and designed and simulated CMOS + FeFET OIM. **b.** Normalized computation time vs logarithm of MIMO size *(log$_2$M)* using the same methods*.* The baseline computation time values for each method (corresponding to 2 × 2 MIMO), which are used for normalization, are mentioned in Table 4. The brute-force ML decoder's computation time is theoretical while that of ZF and OIM are obtained numerically in this work. In (b), of the three methods, only computation time of OIM vs *log$_2$M* plot is linear (blue plot), and linear fitting is carried out as shown through the blue

dotted line. This establishes that OIM exhibits logarithmic time complexity for the MIMO decoding problem.

**Tables and Captions**

| Coefficient | Condition Check | Out-of-Phase Coupling Conductance | In-Phase Coupling Conductance |
|---|---|---|---|
| $a_i$ | Is $a_i > 0$? | $G_{01}^+(\|a_i\|)$ $= \frac{(\|a_i\| - C_{min})}{C_{max} - C_{min}}(G_{max} - G_{min}) + G_{min}$ | $G_{min}$ |
| $a_i$ | Is $a_i < 0$? | $G_{min}$ | $G_{01}^-(\|a_i\|)$ $= \frac{(\|a_i\| - C_{min})}{C_{max} - C_{min}}(G_{max} - G_{min}) + G_{min}$ |
| $b_{ij}$ | Is $b_{ij} > 0$? | $G_{ij}^+(\|b_{ij}\|)$ $= \frac{(\|b_{ij}\| - C_{min})}{C_{max} - C_{min}}(G_{max} - G_{min}) + G_{min}$ | $G_{min}$ |
| $b_{ij}$ | Is $b_{ij} < 0$? | $G_{min}$ | $G_{ij}^-(\|b_{ij}\|)$ $= \frac{(\|b_{ij}\| - C_{min})}{C_{max} - C_{min}}(G_{max} - G_{min}) + G_{min}$ |

**Table 1.** Mapping between coefficients of the maximum likelihood (ML) objective function/ Ising Hamiltonian to be optimized (equation 4) and the coupling conductance

values of the crossbar array in the designed OIM (meanings of all symbols explained in the text)

| Coupling Conductance Range ($G_{min}$ to $G_{max}$) | BER for 10 × 10 MIMO | BER for 20 × 20 MIMO | BER for 40 × 40 MIMO |
|---|---|---|---|
| 1 nS to 100 nS | 0.49 | 0.51 | 0.5325 |
| 10 nS to 1 μS | 0.35 | 0.355 | 0.3625 |
| 100 nS to 10 μS | 0.07 | 0.04 | 0.0475 |
| **1 μS to 100 μS** | **0** | **0** | **0** |
| 10 μS to 1 mS | 0 | 0.005 | 0.1125 |
| 100 μS to 10 mS | No oscillation | No oscillation | No oscillation |
| 1 mS to 100 mS | No oscillation | No oscillation | No oscillation |

**Table 2.** Comparison of bit error rate (BER) for different MIMO sizes as the conductance range of the coupling element in the crossbar array is varied in the SPICE simulation. 10 use cases (as defined in text) are considered for each MIMO size. No oscillation is observed in the OIM circuit when the coupling conductance is too high, and hence no BER values can be reported in these cases. The coupling conductance range of 1 μS to 100 μS turns out to be the most appropriate range, and in the FeFET device we have experimentally demonstrated this conductance range.

| NVM Device | Materials Stack | Coupling Conductance Range ($G_{min}$ to $G_{max}$) | Related Experimental Report |
|---|---|---|---|
| | | | |
| RRAM | Ti/HfO$_2$/HfO$_{2-x}$/ Pt | ≈ 0.5 mS to ≈ 2.5 mS | [67] |
| RRAM | AlO$_x$/HfO2/Ti/TiN | 20 $\mu$S to 60 $\mu$S | [68] |
| RRAM | Al/ALPO/ITO/SiO$_2$/Si | 30 $\mu$S to 70 $\mu$S | [69] |
| RRAM | W/WO$_x$/graphene/Li$_3$PO$_4$/Si/W | ≈ 0.6 $\mu$S to ≈ 1.7 $\mu$S | [70] |
| Spintronic | Ta/CoFeB/MgO | 0.95 mS to 1.1 mS | [71] |
| FeFET | p-Si/SiO$_2$/Hf$_{0.5}$Zr$_{0.5}$O$_2$/TiN | ≈ 1 $\mu$S to ≈ 60 $\mu$S | [31,72] |
| **FeFET** | **p-Si/SiO$_2$/HfO$_2$/metal** | **≈ 1 $\mu$S to ≈ 60 $\mu$S** | **This work** |

**Table 3.** Comparison table of conductance ranges offered by different emerging NVM devices (based on recent experimental reports) showing that only the FeFET device offers the suitable conductance range for obtaining low BER (as identified in Table 2: 1 $\mu$S to 100 $\mu$S approximately) among these devices.

| **MIMO Decoding Method** | **Computation time ($2 \times 2$ MIMO)** |
|---|---|
| Zero-Forcing (ZF) | 12959.0 ns |
| CMOS + FeFET OIM | 20.23 ns |

**Table 4.** Computation time values for 2 × 2 MIMO system (baseline values used for normalization in Fig. 5).

# Supplementary Information for "Symbol Detection in a MIMO Wireless Communication System Using a FeFET-coupled CMOS Ring Oscillator Array"

**S1. SPICE Simulation Results for an Individual Ring Oscillator Block**

The design of an individual CMOS ring oscillator block in our work is shown in Fig. 1b of the main article. Here, we show SPICE simulation results of this block using the BSIM4 transistor model. The method of the simulation and relevant parameters are discussed in the text of the main article. Output voltage at the output terminal of the ring oscillator block, as labelled in Fig. 1b of the main article, is plotted as a function of time in Fig. S1a here. A square-shaped waveform with a fixed time period is obtained as expected from ring oscillators, as also reported elsewhere[R1,R2].

The natural frequency of the oscillation is identified to be 1.24 GHz: main peak in the Fast Fourier Transform (FFT) of the waveform, as shown in Fig. S1b. Several other secondary peaks are also observed in the FFT spectrum because the output voltage waveform in Fig. S1a is a square waveform and not a sinusoidal one.

**S2. SPICE Simulation of OIM for Symbol Decoding in a 3 × 3 MIMO System**

We show an example here of how we have used SPICE simulation of the ring-oscillator-based OIM to solve symbol decoding for a particular use case (as defined in the main article text) and a given MIMO size (we consider 3 × 3 MIMO here). We follow the same technique for other use cases and other MIMO sizes.

Let the actual transmitted vector in this use case (considering BPSK modulation) be: $x^T[1] = 1$, $x^T[2] = 1$, $x^T[3] = -1$. Let the channel matrix coefficient **H** and the noise term **n** (equation 1 of the main text) during signal transmission be such that the Ising Hamiltonian/ objective function corresponding to the maximum likelihood detection method (derived following the method described in Results section of the main text) is as follows:

$$f(x) = -11.3x[1] - 16.9x[2] + 12.6x[3] + 1.9x[1]x[2] - 2.2x[1]x[3] - 6x[2]x[3]$$

(S1)

Following the symbol notations and conductance mapping scheme discussed in the main text, the coupling conductance values for the OIM circuit corresponding to the above Hamiltonian (equation S1) are listed in table ST1. It is to be noted that $G_{min}$ = 1 $\mu$S, $G_{max}$ = 60 $\mu$S as per the experimentally measured conductance modulation in the FeFET device (Fig. 3 of the main text).

The OIM circuit designed with the coupling conductance elements set to the values listed in Table ST1 is then simulated in SPICE using the method described in main text. The final voltage waveforms at the output terminals of the reference oscillator 0 and the main oscillators 1, 2, and 3 are shown in Fig. S2. With the simulation carried out up to 2 $\mu$s, the waveforms for only the final 0.01 $\mu$s have been plotted.

Comparing against the voltage waveform of the reference oscillator 0, relative phases for voltage waveforms of the other oscillators 1, 2, and 3 have been found to be 0°, −4°, and −172° respectively. Further, these phases are binarized (as discussed in the main text) to obtain the final predicted transmission vector from the OIM: $x^{OIM}[1] = 1$, $x^{OIM}[2] = 1$, $x^{OIM}[3] = -1$.

To verify if indeed the predicted vector from the OIM matches with the ground state/ minimum value of the Ising Hamiltonian/ objective function in equation S2 above, all possible combinations of ($[x[1]$, $x[2]$, $x[3]]$) are considered, and the corresponding Hamiltonian/ objective function values (equation S1) are calculated and listed in Table ST2. This is the brute force method for calculating maximum likelihood as described in the main text.

It is evident from Table ST2 that the value of the Hamiltonian / objective function $f(x)$ is the lowest for the vector $x[1] = 1$, $x[2] = 1$, $x[3] = -1$, which matches with the vector predicted from SPICE simulation of the OIM circuit: $x^{OIM}[1] = 1$, $x^{OIM}[2] = 1$, $x^{OIM}[3] = -1$. The predicted vector from the OIM circuit also matches with the actual vector that has been transmitted: $x^T[1] = 1$, $x^T[2] = 1$, $x^T[3] = -1$.

Thus, for this MIMO size and this particular use case, the designed OIM circuit yields the correct solution. We match the predicted vector from OIM with both the actual transmission vector and that which yields the minimum of the maximum likelihood objective function because in some cases, where the noise is too high, even the minimum of the maximum likelihood objective function calculated through brute force (Table ST2) may not match with the actual transmitted vector.

**Supplementary Figures and Captions**

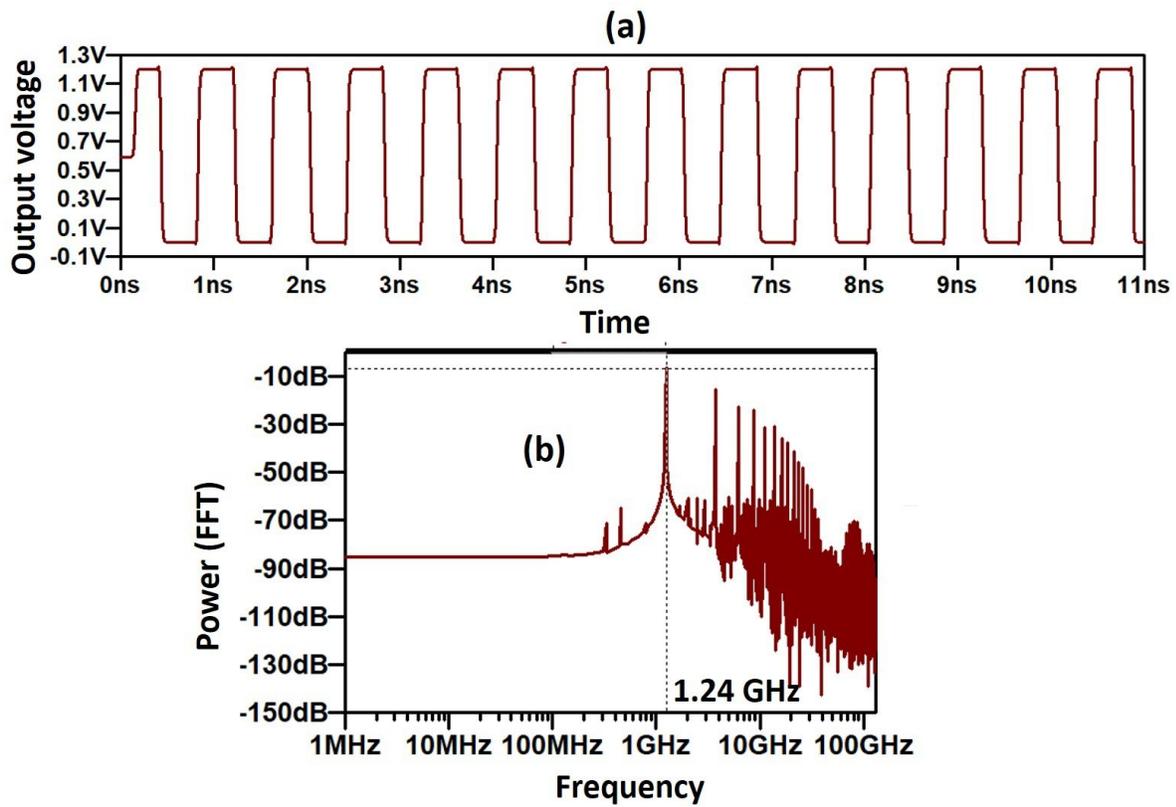

**Fig. S1.** Individual ring oscillator block. **a.** Voltage at the output terminal of the ring oscillator block as a function of time. **b.** FFT of the time-dependent signal in (a)

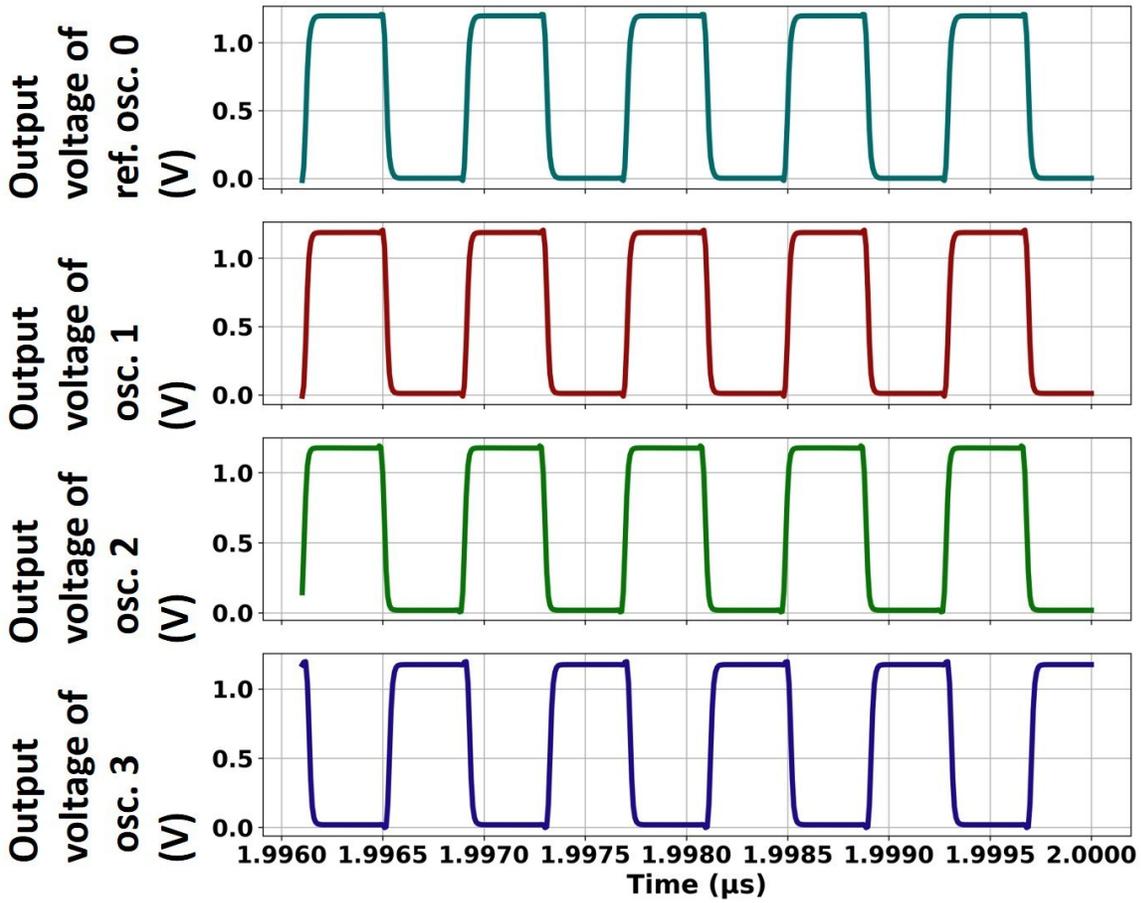

**Fig. S2.** Voltage waveforms at the output terminals of the reference oscillator 0 and three main oscillators 1, 2, and 3 for a transmission case in a 3 × 3 MIMO system

**Supplementary Tables and Captions**

| Coefficient | Conductance Values |
|---|---|
| $a_1 = -11.3$ | $G_{01}^+ = 1\ \mu S,\ G_{01}^- = 40.11\ \mu S$ |
| $a_2 = -16.9$ | $G_{02}^+ = 1\ \mu S,\ G_{02}^- = 60\ \mu S$ |
| $a_3 = 12.6$ | $G_{03}^+ = 44.73\ \mu S,\ G_{03}^- = 1\ \mu S,$ |
| $b_{12} = 1.9$ | $G_{12}^+ = 6.74\ \mu S,\ G_{12}^- = 1\ \mu S, S$ |
| $b_{13}$ = 2.2 μS | $G_{13}^+ = 1\ \mu S,\ G_{13}^- = 7.81\ \mu S,$ |
| $b_{23}$ = -6 μS | $G_{23}^+ = 1\ \mu S,\ G_{23}^- = 21.3\ \mu S,$ |

**Table ST1.** Mapping of coefficients in Ising Hamiltonian to coupling conductance values in the OIM circuit for transmission in a 3 × 3 MIMO system

| $x[1]$ | $x[2]$ | $x[3]$ | $f(x)$ |
|---|---|---|---|
| 1 | 1 | 1 | -21.9 |
| **1** | **1** | **-1** | **-30.7** |
| 1 | -1 | 1 | 20.1 |
| 1 | -1 | -1 | -12.7 |
| -1 | 1 | 1 | 1.3 |
| -1 | 1 | -1 | -16.3 |
| -1 | -1 | 1 | 50.9 |
| -1 | -1 | -1 | 9.3 |

**Table ST2.** Ising Hamiltonian/ objective function values for all possible ($[x[1]$, $x[2]$, $x[3]]$) combinations according to equation S1. The combination which yields the lowest objective function value is shown in bold.